\input harvmac
\noblackbox
\newcount\figno
\figno=0
\def\fig#1#2#3{
\par\begingroup\parindent=0pt\leftskip=1cm\rightskip=1cm\parindent=0pt
\baselineskip=11pt
\global\advance\figno by 1
\midinsert
\epsfxsize=#3
\centerline{\epsfbox{#2}}
\vskip 12pt
\centerline{{\bf Figure \the\figno:} #1}\par
\endinsert\endgroup\par}
\def\figlabel#1{\xdef#1{\the\figno}}

\def\pl#1#2#3{Phys. Lett. {\bf B#1} (#2) #3}


\font\cmss=cmss10
\font\cmsss=cmss10 at 7pt
\def\rlx{\relax\leavevmode}
\def\inbar{\vrule height1.5ex width.4pt depth0pt}
\def\IC{\relax\,\hbox{$\inbar\kern-.3em{\rm C}$}}
\def\IN{\relax{\rm I\kern-.18em N}}
\def\IP{\relax{\rm I\kern-.18em P}}
\def\ZZ{\rlx\leavevmode\ifmmode\mathchoice{\hbox{\cmss Z\kern-.4em Z}}
 {\hbox{\cmss Z\kern-.4em Z}}{\lower.9pt\hbox{\cmsss Z\kern-.36em Z}}
 {\lower1.2pt\hbox{\cmsss Z\kern-.36em Z}}\else{\cmss Z\kern-.4em
 Z}\fi}
\def\IZ{\relax\ifmmode\mathchoice
{\hbox{\cmss Z\kern-.4em Z}}{\hbox{\cmss Z\kern-.4em Z}}
{\lower.9pt\hbox{\cmsss Z\kern-.4em Z}}
{\lower1.2pt\hbox{\cmsss Z\kern-.4em Z}}\else{\cmss Z\kern-.4em
Z}\fi}

\def\narrowplus{\kern -.04truein + \kern -.03truein}
\def\narrowminus{- \kern -.04truein}
\def\narrowminussub{\kern -.02truein - \kern -.01truein}

\def\a{{\alpha}}

\def\r{{\rightarrow}}

\def\frac#1#2{{#1\over #2}}

\def\IZ{\relax\ifmmode\mathchoice
{\hbox{\cmss Z\kern-.4em Z}}{\hbox{\cmss Z\kern-.4em Z}}
{\lower.9pt\hbox{\cmsss Z\kern-.4em Z}}
{\lower1.2pt\hbox{\cmsss Z\kern-.4em Z}}\else{\cmss Z\kern-.4em
Z}\fi}
\def\IB{\relax{\rm I\kern-.18em B}}
\def\IC{{\relax\hbox{$\inbar\kern-.3em{\rm C}$}}}
\def\ID{\relax{\rm I\kern-.18em D}}
\def\IE{\relax{\rm I\kern-.18em E}}
\def\IF{\relax{\rm I\kern-.18em F}}
\def\IG{\relax\hbox{$\inbar\kern-.3em{\rm G}$}}
\def\IGa{\relax\hbox{${\rm I}\kern-.18em\Gamma$}}
\def\IH{\relax{\rm I\kern-.18em H}}
\def\II{\relax{\rm I\kern-.18em I}}
\def\IK{\relax{\rm I\kern-.18em K}}
\def\IP{\relax{\rm I\kern-.18em P}}

\font\cmss=cmss10 \font\cmsss=cmss10 at 7pt
\def\IR{\relax{\rm I\kern-.18em R}}

\def\1{{\bf 1}}
\def\3{{\bf 3}}
\def\7{{\bf 7}}
\def\6{{\bf 6}}
\def\2{{\bf 2}}
\def\8{{\bf 8}}

\def\t{\theta}
\def\hF{\widehat F}

\def\ttheta{ \widetilde \theta}
%

%
%
\def\eqnn#1{\xdef #1{(\secsym\the\meqno)}\writedef{#1\leftbracket#1}%
\global\advance\meqno by1\wrlabeL#1}
\def\eqna#1{\xdef #1##1{\hbox{$(\secsym\the\meqno##1)$}}
\writedef{#1\numbersign1\leftbracket#1{\numbersign1}}%
\global\advance\meqno by1\wrlabeL{#1$\{\}$}}
\def\eqn#1#2{\xdef #1{(\secsym\the\meqno)}\writedef{#1\leftbracket#1}%
\global\advance\meqno by1$$#2\eqno#1\eqlabeL#1$$}

\lref\rCDS{A. Connes, M. R. Douglas and A. Schwarz, hep-th/9711162, JHEP 
9802:003, 1998.}
\lref\rSW{N. Seiberg and E. Witten, hep-th/9908142, JHEP 9909:032, 1999. }
\lref\rMO{C. Montonen and D. Olive, \pl{72}{1977}{117}.}
\lref\rsen{A. Sen, hep-th/9402032, \pl{329}{1994}{217}.}
\lref\rLS{N. Seiberg, L. Susskind and N. Toumbas, hep-th/0005015.}
\lref\rKJ{M. Kreuzer and J.-G. Zhou, hep-th/9912174,  JHEP 0001:011, 2000. }

\Title{\vbox{\hbox{hep-th/0005046}
\hbox{IASSNS--HEP--00/37, PUPT-1929}}}
{\vbox{\centerline{Duality and Non-Commutative Gauge Theory}}}
\centerline{ 
Ori J. Ganor$^\dagger$\footnote{$^1$}{origa@viper.princeton.edu},
Govindan Rajesh$^\ast$\footnote{$^2$}{rajesh@sns.ias.edu} and 
Savdeep Sethi$^{\ast }$\footnote{$^3$}{sethi@sns.ias.edu}}

\vskip 0.1 in
\medskip\centerline{$\ast$ \it School of Natural Sciences, Institute for
Advanced Study, Princeton, NJ 08540, USA}
\medskip\centerline{$\dagger$ \it Department of Physics, Jadwin Hall,
Princeton, NJ 08544, USA}

\vskip 0.5in

We study the generalization of S-duality to non-commutative gauge theories. For rank
one theories, we obtain the leading terms of the dual theory by 
Legendre transforming the Lagrangian of the non-commutative theory 
expressed in terms of a commutative gauge field. 
The dual description is weakly coupled when the original theory is strongly coupled
if we appropriately scale the non-commutativity parameter. 
However, the dual theory appears to be non-commutative in space-time when the 
original theory 
is non-commutative in space. This suggests that locality in time for non-commutative
theories is an artifact of perturbation theory. 

\vskip 0.1in
\Date{5/00}

\newsec{Introduction}

Non-commutative gauge theory \rCDS\ provides an interesting class of examples in 
which to explore the effects of spatial non-locality. While it is easy to define the 
classical non-commutative gauge theory, it is much harder to determine whether the
quantum theory exists. Since non-commutative gauge theories arise in particular
string theory backgrounds, we know that these theories can be embedded consistently
in string theory. The decoupling argument of Seiberg and Witten 
\rSW\ suggests that some of these
theories might exist as quantum theories independent of string theory.  

We are primarily interested in four-dimensional gauge theories. Our goal is to understand
how S-duality \refs{\rMO, \rsen} generalizes to non-commutative gauge theory.
The generalization is not a straightforward consequence of S-duality in type IIB
string theory. To see this, let us begin by briefly recalling how S-duality 
of N=4 Yang-Mills arises from string theory. In the limit $\a'\r 0$, the theory 
on coincident D3-branes is N=4 Yang-Mills. For simplicity, we set the 
RR scalar $C^{(0)}$ to zero. The gauge theory coupling 
constant, $g^2$, is then related to the closed string coupling 
constant $g_s = e^{\phi}$: 
\eqn\trivialrel{ {g^2 \over 4 \pi} =  g_s. }
The conjectured $SL(2, \IZ)$ symmetry of string theory then descends to an $SL(2, \IZ)$
symmetry of the field theory.

To obtain non-commutative Yang-Mills, we consider a system of coincident D3-branes
with NS-NS $B$-field non-zero along the brane. 
In the decoupling limit \rSW, the theory on the brane has a
coupling constant related to the open string coupling constant, $G_s$, rather than
the closed string coupling:
\eqn\noncommcoupling{ g^2 = 2 \pi G_s.}  
In the decoupling limit, the closed string coupling constant goes to zero 
while $G_s$ remains finite and dependent on the $B$-field. 
In this case, S-duality
of the closed string theory does not descend to a symmetry of the field theory. 

For a $U(1)$ gauge theory, S-duality can be demonstrated
directly with a purely field theoretic argument. We start with the Minkowski 
space action,\foot{We use $*$ to denote the Hodge dual of a form rather than the
star product.} 
\eqn\abel{ S = - \int{ \,  {1\over 4 g^2} F \wedge *F },}
where $F = d A$ is the field strength. We want to perform a Legendre transformation
with respect to $F$. To implement the Bianchi identity, 
$$ dF = 0, $$
we introduce a dual gauge-field $A_D$, 
\eqn\bianchi{ S = - \int{ \,  \left( {1\over 4 g^2} F \wedge *F 
+ {1\over 2} A_D \wedge dF \right). }}
We can now treat $F$ as an independent variable and perform the path-integral
over $F$. This amounts to solving the field equations for $F$ which gives the
relation,
\eqn\field{ d A_D = {1\over g^2} \, *F, }
and the resulting dual action,
\eqn\dualaction{  S = - \int{ \,  {g^2\over 4} F_D \wedge * F_D }.}
The aim of this discussion is to generalize this purely field theoretic argument 
to the non-commutative rank one theory. Unlike ordinary abelian gauge theory, the 
coupling constant cannot be scaled away even for the rank one non-commutative theory.

In the following section, we explicitly 
show that the non-commutative action expressed in terms of a commutative gauge-field 
contains only powers of $F$ to order $\theta^2$. In particular, the gauge-field does
not appear explicitly. It is not hard to argue that this must be true to all orders
in $\theta$. This implies that we can obtain a dual description by Legendre transforming
with respect to 
$F$. The resulting dual theory is classical since we neglect loops. However, to order
$\theta$, we will see that no loops appear and the quantum and semi-classical dual 
descriptions agree. To order $\theta^2$, loops appear and the bosonic theory needs to be
regulated. At this point, the computation should be performed in the full N=4 theory.

Fortunately, our primary observations are already visible at order $\theta$. We find 
that under the duality transformation, 
\eqn\result{ \theta \quad  \r  \quad \ttheta  =  g^2 (* \theta). }
That this transformation does not square to one is not so surprising since $(S)^2$ 
is not the identity
operation but charge conjugation. We will also find that $\ttheta$ must be held fixed 
if the dual theory is to have a perturbative expansion in $1/g$.  Even
more interesting is the observation that if $\theta$ is purely spatial then $\ttheta$
involves a space direction and a time direction. The theory becomes non-commutative
in space-time. Although we will not obtain the complete quantum dual description, it
seems clear that this feature, visible at leading order in $\theta$, persists to higher
orders. Space-time non-commutative theories are highly unusual; see \rLS\ for a recent
discussion. Our result suggests that we cannot avoid studying these theories if we
are to understand theories which perturbatively have only spatial non-commutativity.

\newsec{The Duality Transformation}
\subsec{Rewriting the non-commutative Lagrangian}

The non-commutative theory is defined by the action,
\eqn\noncommaction{ S = - {1\over 4 g^2} \int \, \hF \wedge * \hF. }
The change of variables given in \rSW\ allows us to express $\hF$ in terms 
of a commutative gauge-field $A$. We assume that $\theta$ is purely spatial. The
relation takes the form,
\eqn\expand{ \hF  = F + T_\theta (A) + T_{\theta^2}(A) + \ldots.}
The terms of order $\theta$ are given by, 
\eqn\oone{ T_\theta (A) =  - F \t F - A_k \t^{kl} \partial_l F. }
We follow the notation of \rKJ\ where $ F \t F = F_{ik} \t^{kl} F_{kj}$.   
The expression for $T_{\theta^2}(A)$ is found in \rKJ, 
\eqn\otwo{ \eqalign{T_{\theta^2}(A) = & F \t F \t F + {1\over 2} A_k \t^{kl} 
\left( \partial_l
A_m  + F_{lm} \right) \t^{mn} \partial_n F \cr
& + \t^{kl} A_k \partial_l \left( F\t F\right)
+ {1\over 2} \t^{kl} \t^{mn} A_k A_m \partial_l \partial_n F.\cr }}
The expression for $\hF$ explicitly contains $A$. However, we can manipulate the
action \noncommaction\ so that it takes the following form,
\eqn\expandaction{ S =  - {1\over 4 g^2} \int \,  \left( 
F \wedge *F + L_\t (F) + L_{\t^2}(F) + \ldots \right).}
The terms of order $\theta$ take the form,
\eqn\lone{ L_\t (F) =  2 \tr ( \t F^3) - {1\over 2} \tr (\t F) \, \tr (F^2),}
where we define $\tr (AB) = A_{ij} B^{ji}. $ Since our theory is rank one, there
should be no confusion with traces over group indices. It is not too hard to 
find an expression for $ L_{\t^2}(F)$ which takes the form:
\eqn\ltwo{ \eqalign{ L_{\t^2}(F) = &  - 2 \tr (\t F \t F^3) + \tr (\t F^2 \t F^2) 
+ \tr (\t F) \, \tr ( \t F^3) \cr & - {1\over 8} 
\tr (\t F)^2 \, \tr (F^2) + {1\over 4} \tr (\t F \t F) \, \tr (F^2). \cr }}
While we have explicitly demonstrated that it is possible to express  
\noncommaction\ in terms of $F$ to order $\theta^2$, it must be the case to all 
orders in $\theta$. The only gauge-invariant operator that can be constructed from
$A$ is $F$. While $\hF$ can depend on $A$ explicitly, the action must be gauge-invariant
under the commutative gauge-invariance. This requires that the action be expressible
in terms of $F$ alone. 

\subsec{Duality at $O(\theta)$}

Since the action can be expressed in terms of $F$, we can implement a duality
transformation in essentially the way described in the introduction.  
To perform the Legendre transform, we shift the action as before
\eqn\shift{ S \quad \r \quad S + \int {1\over 2} A_D \wedge dF. }
The equation of motion for $F$ gives, 
\eqn\eom{ g^2 F_D =   *F  + {1\over 2} {\delta L_{\t} \over \delta F}(F) + O(\t^2). }
To lowest order in $\t$, we can solve for $F$ in terms of $F_D$:
\eqn\solveF{ *F =  g^2 F_D - {1\over 2} {\delta L_{\t} \over \delta F} 
\bigg\vert_{F=- * g^2 F_D}
+  O(\t^2).}
At order $\theta$, loops play no role in the duality transformation so the quantum 
and semi-classical dual descriptions are equivalent. Plugging \solveF\ into the action
\expandaction\ gives, 
\eqn\dualform{ S = - {g^2\over 4} \int \,   \left(
F_D \wedge *F_D  +  2 \tr ( \ttheta F_D^3) - {1\over 2} \tr (\ttheta F_D) \, 
\tr (F_D^2) \right)+ O(\ttheta^2). }
Note that we use $\ttheta = g^2 (* \theta)$ as the new non-commutativity parameter.
The factor of $g^2$ in $\ttheta$ 
is natural because of the following scaling argument: we can
schematically expand $\hF^2$, 
\eqn\dimexp{ \hF^2 \sim  F^2 \left( 1 + \sum_{n,l} \theta^{n+l} (\partial )^{2l} F^n
\right),}
on strictly dimensional grounds. This implies that iteratively, we can express
$F$ in schematic form:
\eqn\fschem{ F \sim - g^2 * F_D \left( 1 + \sum _{n,l} \theta^{n+l} 
(\partial )^{2l} ( g^2 *F_D)^n  \right).} 
In terms of $\ttheta$, we see that
\eqn\reschem{ F \sim - g^2 * F_D \left( 1 + \sum _{n,l} \ttheta^{n+l} 
(\partial )^{2l} \left( { 1\over g^2} \right)^{l} ( *F_D)^n  \right).} 
The action now takes the form of a derivative expansion with higher derivatives of $F_D$
suppressed by powers of $ g^{-1}$. 

There are a number of observations at this point. Substituting even the lowest 
order expression,
\eqn\low{ F = - g^2 *F_D + O(\t), }
into \expandaction\ results in an infinite
number of terms involving higher powers of $\ttheta$. While terms beyond $O(\ttheta)$
will receive additional corrections from the $O(\t)$ corrections to \low, it seems 
quite clear -- barring 
miraculous cancellations -- that there is no upper bound on the power of $\ttheta$ that 
appears in the dual action. This suggests that it will be difficult to
quantize the theory  non-perturbatively in any conventional way. 
We also note that the dual action to leading order in $\ttheta$, 
expressed in dual non-commutative variables, takes the form
\eqn\dualaction{ S = - {g^2\over 4} \int  \hF_D \wedge \hF_D + O(\ttheta^2).}
As is natural, we define $\hF_D$ with respect to a star product involving $\ttheta$. 
However, it is quite possible that the corrections to \dualaction\ of $O(\ttheta^2)$
are non-vanishing. It is not clear that the resulting dual action would then 
have a purely quadratic form.

\bigbreak\bigskip\bigskip\centerline{{\bf Acknowledgements}}\nobreak

It is our pleasure to thank M. R. Douglas and N. Seiberg for helpful comments,
and K. Dasgupta for early participation.
The work of O.J.G. is supported in part by NSF Grant No. PHY-98-02484. 
The work of R.G. is supported in part
by NSF grant No. DMS-9627351, while that of 
S.S. is supported  by the William Keck Foundation and by 
NSF grant No. PHY--9513835.

\vfill\eject

\listrefs
\bye